\documentclass[12pt]{report}
\usepackage[utf8]{inputenc}
\usepackage{blindtext}
\usepackage[letterpaper, total={8.5in, 11in}, margin=2in]{geometry}
\usepackage{ragged2e}
\usepackage{blindtext}
\usepackage{graphicx}
\usepackage{amsmath}
\usepackage{amssymb}
\usepackage{gensymb}
\usepackage{array}
\date{}

\begin{document}
\centering{\huge Intelligent Reflecting Surfaces for the Enhancement of 6G Internet of Things\\
\vspace{24pt}
\large Mobasshir Mahbub, Raed M. Shubair}

\newpage

\RaggedRight{\textbf{\Large 1.\hspace{10pt} Introduction}}\\
\vspace{18pt}
\justifying{\noindent The last three decades have witnessed an exponential growth and tremendous developments in wireless technologies and techniques, and their associated applications.  These include indoor localization techniques and related aspects [1-39], terahertz communications and signal processing applications [40-65], and antenna design and propagation characteristics [66-100].

Mobile and wireless connectivity have played a crucial role in the prevailing economy [101] with technological advances that reasonably satisfy the subscriber end with substantial involvement in business, education, transportation, and other principal industrial functionalities. These are now capable of connecting people and devices for unprecedented exchanges of audiovisual and data content, experiencing their fastest development in history thanks to enabling technologies for promoting widespread adoption and further enhancing connectivity [102].

According to Cisco, there is a prediction of the influence of visual networking technologies on global networks with a projected spike of 77 exabytes in 2023 increased from 11.5 exabytes in 2017. In the future years, the compound annual growth rate (CAGR) is predicted to be increased around 74 percent of current mobile data traffic [103].

With the advancement of sensing technologies over the years, it has become critical to ensure the seamless connectivity of the Internet of Things (IoT) gadgets. With the advancement of communication technology, cellular networks are increasingly being utilized to link IoT systems. According to the concept of IoT, “things” or “objects” on the planet will have a distinctive (electronic) sign which can be recognized by other objects, and every electronic object or thing should be interconnected [104]. Furthermore, information may be transmitted between objects autonomously. The number of things linked to the network will be far more than the number of the population. The objects or things will be the primary source of data transmission. The current human-to-human connectivity will be replaced by the interaction between human-to-object and object-to-object. The future connectivity trend is not to be interpersonal communications, and the information will not be obtained from a human, but rather from the interaction between vast numbers of things that can represent human [105]. The real world or the actual environment may be integrated with the virtual ecosystem of information via the IoT [106]. The cognition of the physical world is a crucial component of it.

IoT is evolving at an unprecedented rate, driven by tremendous demand in a variety of applications. According to a recent prediction by the International Data Corporation (IDC), 41.6 billion linked IoT products are to be performing (operating) in 2025, creating 79.4 zettabytes of data [107].

The constant advancement of the 5G connectivity has empowered the IoT [108] to link more items, allowing more implementation spaces in a variety of sectors. 5G infrastructure has steadily improved [109]. Although fifth-generation (5G) [10] and beyond 5G (B5G) [111] technologies are still to be adopted globally, the problems of offering lower data rates and limited capacity have prompted researchers toward sixth-generation (6G) technologies [111], [112], [113]. Developed countries and advanced research organizations have begun to study and construct 6G technologies for next-generation network technologies. The developing mobile content in this context will operate in a frequency range higher than 5G technology. As a result, the associated network connectivity will be greater, the network throughput will be higher, and the overall energy usage (consumption) should be lower.

The sixth generation of wireless communication systems aims to create a data-driven, sustainable future that is facilitated by near-instant, secure, limitless, and green communication [114]. To achieve such a goal, industry and academia have set rigorous performance standards in terms of safety and trust, speed, sensing capabilities, reliability, adaptability, and energy consumption. As a result, the 6G or sixth-gen cellular IoT infrastructure must be designed to fulfill increased needs and expectations, such as broader coverage, better capacities, ubiquitous connectivity, etc. [115]. Therefore, 6G cellular networks have to integrate THz communication [116], [117], Holographic multiple input multiple output (HMIMO) [118], intelligent reflecting surfaces (IRS) [119], [120], symbiotic radio [121], cell-free network access [122], multi-tier heterogeneous network [123], etc. Moreover, it is expected that 6G will incorporate an artificial intelligence-based cloud-edge-device collaborating ground-air-space integrated environment [124].

An IRS [125] is a rectangular metasurface made up of a vast number of reflecting components that has recently gained research attention due to its ability to significantly improve the energy and spectral efficiencies of communication networks by modifying wireless transmission environments. IRS elements can reflect the incoming signal with the required phase shift [126]. IRS creates constructive signal combination and destructive interference suppression at the receivers by dynamically altering the transmissions of the reflected signal. As a result, an improved quality of service (QoS) can be attained at the receiver end.

Deployment of small cells i.e. micro cells [127] in an incredibly crowded network is one of the convenient strategies nowadays for fulfilling the requirements of massive data traffic. Such base stations are installed on premises that are close to the resource-hungry devices usually a shorter indoor/outdoor space to enhance cellular services. Furthermore, such types of base stations have better coverage that guarantees sufficient transmission. As the micro cells are situated close to the users, comparatively a low-power (compared to macro base stations) transmission capacity can assure a higher throughput to the devices [128].

6G cellular IoT must offer precise computation and efficient connectivity for a wide range of devices or objects to allow real-time processing or analysis of large datasets generated by terminal gadgets, which are regarded as two fundamental objectives of cellular IoT in 6G.

IoT services and applications such as smart manufacturing [129], [130], eHealthcare [131], [132], smart cities [133], [134], intelligent home automation [135], smart farming [136], [137], smart grid maintenance and monitoring [138], etc. will flourish and spread highly in the 6G networks because of the enhanced coverage, reliability, extremely low end-to-end delay, etc. Artificial intelligence (AI) [139], machine learning (ML) [140], [141], data analytics i.e. video analytics [142], [143], [144], [145] (e.g. video-processing for vegetation health analysis in smart farming [146], remote monitoring, security surveillance, etc.) will be highly employed in IoT services and applications to strengthen the features of IoT. Therefore, enhanced network coverage and extremely low end-to-end transmission delay should be assured in the context of a 6G [147] IoT or 6G cellular IoT environment.

Xie et al. [148] investigated the weighted sum-rate maximization approach or problem of the IRS-aided multiple-input multiple-output (MIMO) transmission-based IoT network consisting of multiple low-power IoT devices. To perform the analysis, the work developed a joint optimization approach that decomposes the problem into several sub-problems that can be disposed of alternately. Yu et al. [149] designed an IRS-aided massive network framework incorporating a channel estimation process. The work analyzed the performance of the framework. The work derived that IRS escalates the performance of end-users enhancing the network coverage. Mahmoud et al. [150] investigated the deployment of IRS in unmanned aerial vehicles (UAV) enabled wireless communications aiming to enhance the network coverage and escalate the communication consistency as well as spectral efficiency of IoT. Specifically, the research derived analytic expressions for the ergodic capacity, symbol error rate (SER), and outage probability and analyzed the mentioned expressions in the context of the considered network setup. Wu et al. [151] studied the power control in the uplink for IRS-assisted IoT networks under the constraints of quality of service (QoS) requirements at the user end. The research targeted to minimize the total power consumption of the users by jointly optimizing the phase shifts of the reflecting elements of IRS and the receiver beamforming by the base station, considering the signal-to-interference-plus-noise ratio (SINR) constraint for the users. Hao et al. [152] considered an IRS-assisted multiuser MIMO system operating in the THz band with orthogonal frequency-division multiple access (OFDMA) technique. The work adopted the sparse radio frequency (RF) antenna architecture to reduce the power consumption of the network. The target of the research is to maximize the sum rate through the joint optimization of hybrid beamforming i.e. analog/digital at the base station and reflection matrix at the IRS. Yu et al. [153] designed a framework for the IRS-assisted 6G cellular IoT incorporating a channel estimation approach and downlink-uplink data transmission. The work analyzed the performance of the framework to derive the significance of the variation of base parameters of the IRS on spectral efficiency. Chu et al. [154] deployed IRSs to improve the computational efficiency of the multi-access edge computing (MEC) systems for the IoT ecosystem by calibrating the phase shift of the reflecting elements of IRS intelligently. Li et al. [155] utilized the IRS to improve the performance of spectrum detection in IoT devices to minimize energy consumption. Chu et al. [156] analyzed the performance of IRS-assisted wireless powered IoT networks. The research targeted maximizing the sum throughput of the system under the constraints of transmission time scheduling, phase shifts of the elements of IRS, and bandwidth allocation. Okogbaa et al. [157] reviewed the design, implementations, and limitations of the deployment of the IRS in terms of IoT paradigm considering the forthcoming 6G infrastructure. Guo et al. [158] presented a survey on the 6G-empowered massive IoT ecosystem focusing on the drivers, core technical requirements, deployment cases, and trends. However, the work has not described the deployment of IRS as an influencer for IoT services. Verma et al. [159] addressed the issue of green communication for 6G massive IoT networks considering the cluster-based data distribution. The research as well has not considered the deployment of IRS in this circumstance.

Since IRS-assisted communication is an emerging topic and research is ongoing, the authors find that there is a limitation of relative literature during the literature review. According to the literature review, less than 70 articles are currently present in the scholarly research databases that performed analysis on the deployment of IRS for IoT services and applications. Moreover, the literature relative to the considered research concept is unfortunately insufficient (since it’s an emerging topic). Therefore, the review includes literature and works that are nearly relative to the IRS-assisted transmission for IoT.

The paper measured the performance of a conventional micro cellular communication with an IRS-enhanced micro cellular communication considering a two-tier 6G network consisting of a micro cell tier (or layer) operating under a macro cell tier in terms of downlink and uplink received power, signal-to-interference plus noise ratio (SINR), throughput, spectral efficiency, and uplink end-to-end delay (since uplink is the major concern in the case of IoT services and applications) in the context of IoT paradigm. Moreover, the paper analyzed the performance of the probability of user association and the average number of devices considering both of the network scenarios. The research considered passive IRS.

Section 2 includes the measurement model that incorporates several measurement equations. The measurement results and discussions are included in section 3. The paper concludes with section 4 briefing an overview of the work.
}
\vspace{18pt}

\RaggedRight{\textbf{\Large 2.\hspace{10pt} Measurement Model}}\\
\vspace{12pt}

\justifying{\noindent Contemplating a two-tier network consisting of $N$ sets of micro cell base stations (BSs) where $n$is the serving base station serving the $U$ sets of IoT user equipment (micro cell base stations are operating under a macro cell base station). In the case of an IRS-assisted network, the mentioned sets of the base stations will serve the mentioned sets of IoT devices incorporating an IRS in between the base station and user equipment. $P_t^D$ and $B^D$ are the downlink transmit power and bandwidth, respectively. $P_t^U$ and $B^U$ denote the uplink transmit power and bandwidth, respectively. Fig. 1 (a) shows the conventional and IRS-assisted micro cellular network and (b) shows the IRS-assisted micro cellular network for IoT services.}

\begin{figure}[h]
    \centering
    \includegraphics[height=7.5cm, width=10.5cm]{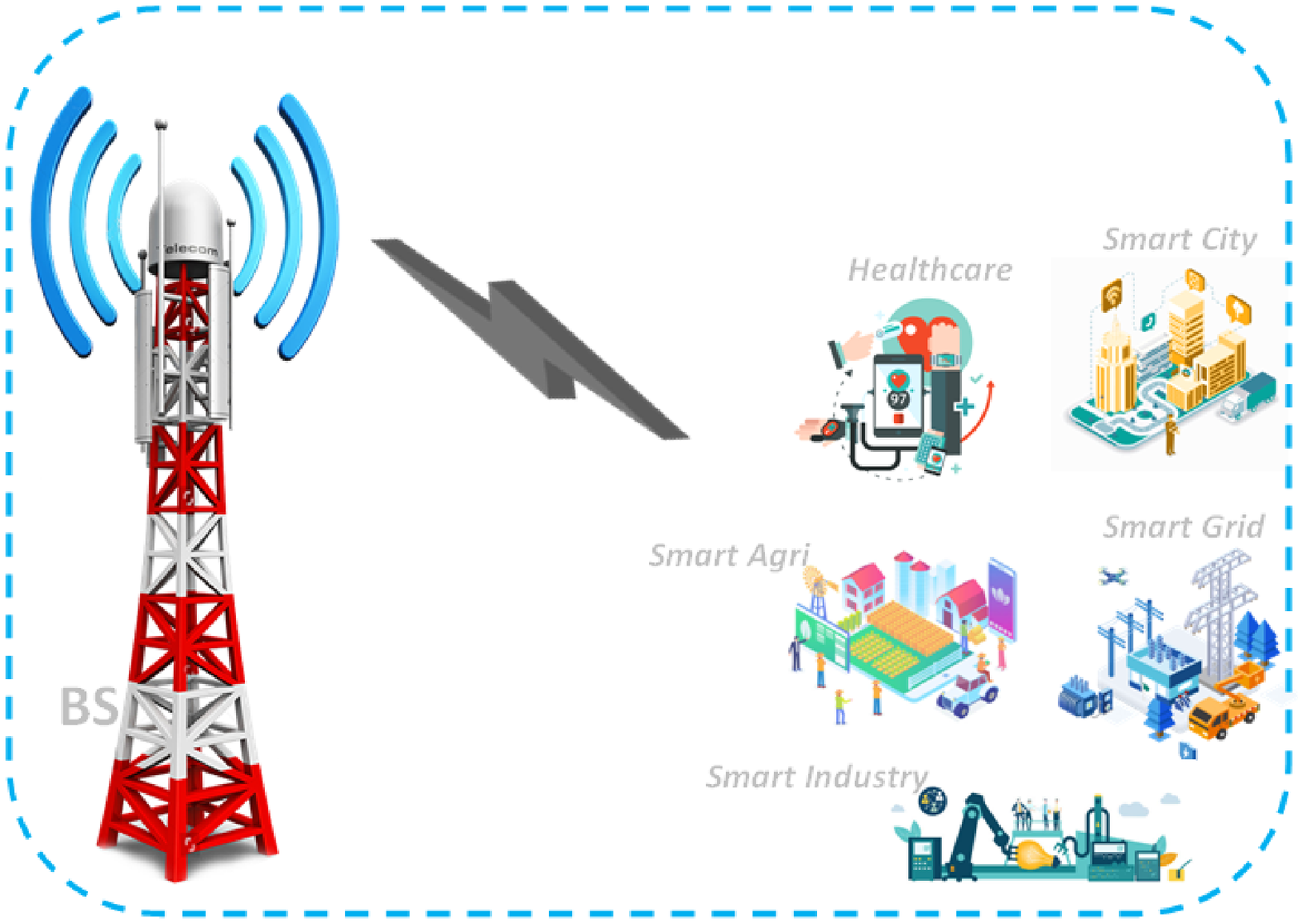}\\
    (a)
\end{figure}

\begin{figure}[h]
    \centering
    \includegraphics[height=7.5cm, width=10.5cm]{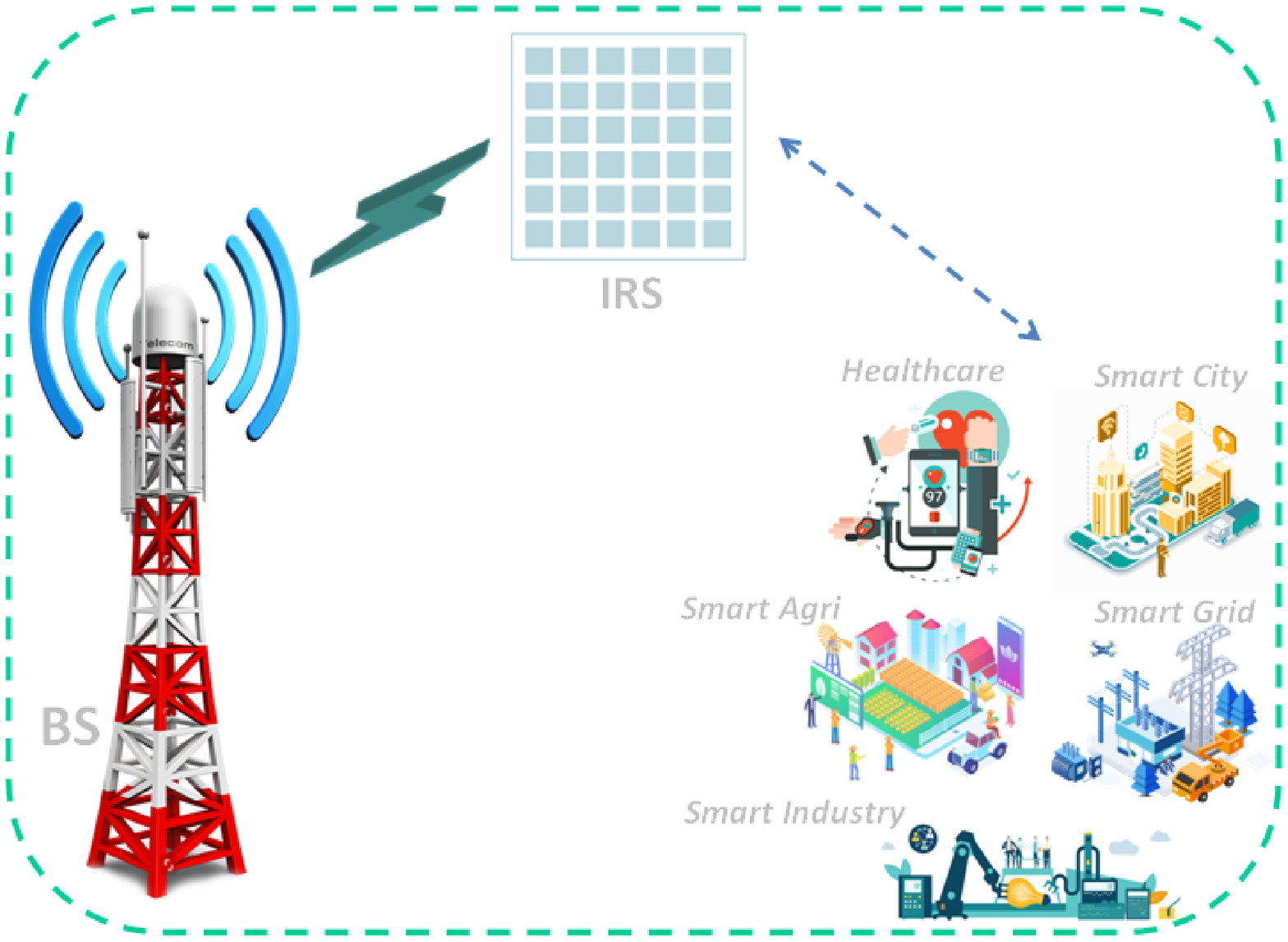}\\
    (b)\\
    
    Fig. 1. (a) Conventional micro cellular network and (b) IRS-assisted micro cellular network
\end{figure}

\vspace{12pt}
\RaggedRight{\textit{\large A.\hspace{10pt} Conventional Micro Cell Model}}\\
\vspace{12pt}
\justifying{\noindent The downlink received power in the case of a conventional micro cell base station is modeled as follows (Eq. 1) [160], [161], [162],
}
\vspace{6pt}
\begin{equation}
P_r^{D}= \frac{P_t^{D}\lambda h}{(4\pi)^2 d^\alpha} 
\end{equation}
where $\lambda= c⁄f_c$  is the wavelength of the carrier. $c$ is the velocity of the light in $ms^{-1}$. $f_c$ denotes the carrier frequency in Hz. $h$ values unit mean exponential distribution that indicated as Rayleigh fading coefficient.

$d= \sqrt{(x^{B}-x^{U})^2+(y^{B}-y^{U})^2+(z^{B}-z^{U})^2}$. $\alpha$ is the exponent representing the degree of attenuation. $P_r^D$ is stated as $P_r^{D(Micro-Conv.)}$ in the case of the probability of user association measurement (in Eq. 19). The uplink received power is measured by (Eq. 2),
\vspace{6pt}
\begin{equation}
P_r^{U}= \frac{P_t^{U}\lambda h}{(4\pi)^2 d^\alpha} 
\end{equation}

The downlink signal-to-interference plus noise ratio (SINR) is calculated by the equation below (Eq. 3),
\vspace{6pt}
\begin{equation}
S^{D}= \frac{P_r^{D}}{I+N} 
\end{equation}
where $I$ is the interference. $N$ is indicating the Gaussian noise.

The SINR for uplink is measured by (Eq. 4),
\vspace{6pt}
\begin{equation}
S^{U}= \frac{P_r^{U}}{I+N} 
\end{equation}

The downlink throughput is formulated by the following equation (Eq. 5) [162]-[166],

\begin{equation}
T^{D}= B^{D}log_2(1+\frac{P_r^{D}}{i+N})
\end{equation}

The throughput in uplink can be determined by (Eq. 6) [162]-[166],

\begin{equation}
T^{U}= B^{U}log_2(1+\frac{P_r^{U}}{i+N})
\end{equation}

The spectral efficiency in downlink is measured by the following equation (Eq. 7),

\begin{equation}
SE^{D}= \frac{T^{D}}{B^{D}} 
\end{equation}

The uplink spectral efficiency is obtained by the equation below (Eq. 8),

\begin{equation}
SE^{U}= \frac{T^{U}}{B^{U}} 
\end{equation}

The transmission delay in uplink is determined by (Eq. 9),

\begin{equation}
d^{U}= \frac{Data}{T^{U}} 
\end{equation}

\vspace{12pt}
\RaggedRight{\textit{\large B.\hspace{10pt} IRS-Assisted Micro Cell Model}}\\
\vspace{12pt}

\justifying{\noindent The received power in the downlink in the case of IRS-assisted micro cellular communication for IoT devices is calculated as follows (Eq. 10) [167],}

\begin{equation}
P_r^{D(IRS)} = \frac{P_t^{D(IRS)}G_t G_r G M^2 N^2 d_x d_y \lambda^2 cos(\theta_t) cos(\theta_r) A^2}{64\pi^3(d_1 d_2)^2}
\end{equation}
where $d_1=  \sqrt{(x^{B}-x_i^{I} )^2+(y^{B}-y_i^{I} )^2+(z^{B}-z_i^{I} )^2}$ denotes the distance between the transmitter (base station) positioned at $(x^{B},y^{B},z^{B})$ coordinates and IRS located at $(x_i^{I},y_i^{I},z_i^{I})$ coordinates. $d_2=  \sqrt{(x_i^{IRS}-x_n^{U} )^2+(y_i^{IRS}-y_n^{U} )^2+(z_i^{IRS}-z_n^{U} )^2}$
is the distance between the IRS and the IoT device (IoTD) located at $(x^{U},y^{U},z^{U})$. $G_t$ is the transmitter gain and $G_r$ is the receiver gain. $G=\frac{4\pi d_x d_y}{\lambda^2}$  represents the scattering gain of the IRS. $M$ and $N$ are the numbers of transmit-receive elements. $d_x$ and $d_y$ denote the length and width of elements. $\lambda$ is the wavelength. $\theta_t$ is the transmit angle and $\theta_r$ is the receive angle. $A$ is the reflection coefficient.

The uplink received power is formulated by the following equation (Eq. 11),

\begin{equation}
P_r^{U(IRS)} = \frac{P_t^{U(IRS)}G_t G_r G M^2 N^2 d_x d_y \lambda^2 cos(\theta_t) cos(\theta_r) A^2}{64\pi^3(d_1 d_2)^2}
\end{equation}

The downlink SINR is obtained by the following equation (Eq. 12),

\vspace{6pt}
\begin{equation}
S^{D(IRS)}= \frac{P_r^{D(IRS)}}{I+N} 
\end{equation}

The SINR in uplink is determined by (Eq. 13),

\vspace{6pt}
\begin{equation}
S^{U(IRS)}= \frac{P_r^{U(IRS)}}{I+N} 
\end{equation}

The downlink throughput is measured by the equation below (Eq. 14) [162]-[166],

\begin{equation}
T^{D(IRS)}= B^{D(IRS)}log_2(1+\frac{P_r^{D(IRS)}}{i+N})
\end{equation}

The uplink throughput is obtained by (Eq. 15) [162]-[166],

\begin{equation}
T^{U(IRS)}= B^{U(IRS)}log_2(1+\frac{P_r^{U(IRS)}}{i+N})
\end{equation}

The downlink spectral efficiency is derived by the following equation (Eq. 16),

\begin{equation}
SE^{D(IRS)}= \frac{T^{D(IRS)}}{B^{D(IRS)}} 
\end{equation}

The uplink spectral efficiency is measured by (Eq. 17),

\begin{equation}
SE^{U(IRS)}= \frac{T^{U(IRS)}}{B^{U(IRS)}} 
\end{equation}

The transmission delay in uplink is derived by the equation below (Eq. 18),

\begin{equation}
d^{U(IRS)}= \frac{Data}{T^{U(IRS)}} 
\end{equation}

In the case of the IRS-assisted micro cellular communication model, the work considered that the network serves each user through the IRS (supported or fed by the micro base station).

\vspace{12pt}
\RaggedRight{\textit{\large C.\hspace{10pt} User Association and Average Number of Devices}}\\
\vspace{12pt}
\justifying
\noindent{This sub-section includes the measurement formulas to analyze and compare the coverage in the case of both conventional micro cell base stations and IRS –assisted micro cell base stations.}

Conventional User Association Model: The user association model for the conventional non-IRS micro cell base station operating under a macro cell is expressed as (Eq. 19) [168],

\begin{equation}
A^{non-IRS} = \left (1+\frac{\lambda_{Mac.}}{\lambda_{Mic.}}\left(\frac{P^{Mac.}_r}{P^{D(Mic.-Conv.}_r}\right)^\frac{2}{\alpha_{mac.}}\right)^{-1}
\end{equation}
where $\lambda_Mac.$ and $\lambda_Mac.$ are the densities (per $m^2$) of the macro and micro cell tier base station/s, respectively. $P_r^{D(Mac.)}$ and $P_r^{D(Mic.-Conv.)}$ indicate the received power from the macro cell base station and conventional micro cell base station, respectively. $\alpha_{mac.}$ denotes the path loss exponent for the macro tier.

IRS-Assisted User Association Model: The user association model for the IRS-assisted micro cellular communication is given by the following equation (Eq. 20) [168],

\begin{equation}
A^{IRS} = \left (1+\frac{\lambda_{Mac.}}{\lambda_{Mic.}}\left(\frac{P^{Mac.}_r}{P^{D(Mic.-IRS)}_r}\right)^\frac{2}{\alpha_{mac.}}\right)^{-1}
\end{equation}
where $P_r^{D(Mic.-IRS)}$ is the received power from the IRS-enhanced micro cell base station.

Number of Associated Devices: The average number of the associated or served devices in the case of conventional micro cellular communication is derived by the equation below (Eq. 21) [169-180],

\begin{equation}
n^{non-IRS} = 1+\left (1.28\lambda_u \frac{\bar{A}^{non-IRS}}{\lambda_{Mic.}}\right)
\end{equation}
where $\lambda_u$ is the user (IoTD) density per $m^2$. $\bar{A}^{non-IRS}$ denotes the averaged probability of user association corresponding to the coverage region in the case of conventional micro cell base station.

The average number of the associated or served devices in the case of IRS-assisted micro cell base station is determined by (Eq. 22),

\begin{equation}
n^{IRS} = 1+\left (1.28\lambda_u \frac{\bar{A}^{IRS}}{\lambda_{Mic.}}\right)
\end{equation}
where $\bar{A}^{IRS}$S denotes the averaged probability of user association in the case of IRS-enhanced micro cell base station.

\vspace{18pt}

\RaggedRight{\textbf{\Large References}}\\

\justifying{
1.	Mohamed I. AlHajri, Nazar T. Ali, and Raed M. Shubair. "Classification of indoor environments for IoT applications: A machine learning approach." IEEE Antennas and Wireless Propagation Letters 17, no. 12 (2018): 2164-2168.

2.	M. I. AlHajri, A. Goian, M. Darweesh, R. AlMemari, R. M. Shubair, L. Weruaga, and A. R. Kulaib. "Hybrid RSS-DOA technique for enhanced WSN localization in a correlated environment." In 2015 International Conference on Information and Communication Technology Research (ICTRC), pp. 238-241. IEEE, 2015.

3.	Mohamed I. AlHajri, Nazar T. Ali, and Raed M. Shubair. "Indoor localization for IoT using adaptive feature selection: A cascaded machine learning approach." IEEE Antennas and Wireless Propagation Letters 18, no. 11 (2019): 2306-2310.

4.	M. A. Al-Nuaimi, R. M. Shubair, and K. O. Al-Midfa. "Direction of arrival estimation in wireless mobile communications using minimum variance distortionless response." In The Second International Conference on Innovations in Information Technology (IIT’05), pp. 1-5. 2005.

5.	Fahad Belhoul, Raed M. Shubair, and Mohammed E. Al-Mualla. "Modelling and performance analysis of DOA estimation in adaptive signal processing arrays." In ICECS, pp. 340-343. 2003.

6.	Ali Hakam, Raed M. Shubair, and Ehab Salahat. "Enhanced DOA estimation algorithms using MVDR and MUSIC." In 2013 International Conference on Current Trends in Information Technology (CTIT), pp. 172-176. IEEE, 2013.

7.	Ebrahim M. Al-Ardi, Raed M. Shubair, and Mohammed E. Al-Mualla. "Direction of arrival estimation in a multipath environment: An overview and a new contribution." Applied Computational Electromagnetics Society Journal 21, no. 3 (2006): 226.

8.	R. M. Shubair. "Robust adaptive beamforming using LMS algorithm with SMI initialization." In 2005 IEEE Antennas and Propagation Society International Symposium, vol. 4, pp. 2-5. IEEE, 2005.

9.	R. M. Shubair and A. Al-Merri. "Robust algorithms for direction finding and adaptive beamforming: performance and optimization." In The 2004 47th Midwest Symposium on Circuits and Systems, 2004. MWSCAS'04., vol. 2, pp. II-II. IEEE, 2004.

10.	Pradeep Kumar Singh, Bharat K. Bhargava, Marcin Paprzycki, Narottam Chand Kaushal, and Wei-Chiang Hong, eds. Handbook of wireless sensor networks: issues and challenges in current Scenario's. Vol. 1132. Berlin/Heidelberg, Germany: Springer, 2020.

11.	R. M. Shubair and Y. L. Chow. "A closed-form solution of vertical dipole antennas above a dielectric half-space." IEEE transactions on antennas and propagation 41, no. 12 (1993): 1737-1741.

12.	Ebrahim M. Al-Ardi, Raed M. Shubair, and Mohammed E. Al-Mualla. "Computationally efficient DOA estimation in a multipath environment using covariance differencing and iterative spatial smoothing." In 2005 IEEE International Symposium on Circuits and Systems, pp. 3805-3808. IEEE, 2005.

13.	R. M. Shubair and W. Jessmi. "Performance analysis of SMI adaptive beamforming arrays for smart antenna systems." In 2005 IEEE Antennas and Propagation Society International Symposium, vol. 1, pp. 311-314. IEEE, 2005.

14.	E. M. Al-Ardi, R. M. Shubair, and M. E. Al-Mualla. "Investigation of high-resolution DOA estimation algorithms for optimal performance of smart antenna systems." (2003): 460-464.

15.	E. M. Al-Ardi, Raed M. Shubair, and M. E. Al-Mualla. "Performance evaluation of direction finding algorithms for adapative antenna arrays." In 10th IEEE International Conference on Electronics, Circuits and Systems, 2003. ICECS 2003. Proceedings of the 2003, vol. 2, pp. 735-738. IEEE, 2003.

16.	M. I. AlHajri, N. Alsindi, N. T. Ali, and R. M. Shubair. "Classification of indoor environments based on spatial correlation of RF channel fingerprints." In 2016 IEEE international symposium on antennas and propagation (APSURSI), pp. 1447-1448. IEEE, 2016.

17.	Mohamed AlHajri, Abdulrahman Goian, Muna Darweesh, Rashid AlMemari, Raed Shubair, Luis Weruaga, and Ahmed AlTunaiji. "Accurate and robust localization techniques for wireless sensor networks." arXiv preprint arXiv:1806.05765 (2018).

18.	Goian, Mohamed I. AlHajri, Raed M. Shubair, Luis Weruaga, Ahmed Rashed Kulaib, R. AlMemari, and Muna Darweesh. "Fast detection of coherent signals using pre-conditioned root-MUSIC based on Toeplitz matrix reconstruction." In 2015 IEEE 11th International Conference on Wireless and Mobile Computing, Networking and Communications (WiMob), pp. 168-174. IEEE, 2015.

19.	Raed M. Shubair, and Ali Hakam. "Adaptive beamforming using variable step-size LMS algorithm with novel ULA array configuration." In 2013 15th IEEE International Conference on Communication Technology, pp. 650-654. IEEE, 2013.

20.	R. M. Shubair, and A. Merri. "Convergence of adaptive beamforming algorithms for wireless communications." In Proc. IEEE and IFIP International Conference on Wireless and Optical Communications Networks, pp. 6-8. 2005.

21.	Zhenghua Chen, Mohamed I. AlHajri, Min Wu, Nazar T. Ali, and Raed M. Shubair. "A novel real-time deep learning approach for indoor localization based on RF environment identification." IEEE Sensors Letters 4, no. 6 (2020): 1-4.

22.	R. M. Shubair, A. Merri, and W. Jessmi. "Improved adaptive beamforming using a hybrid LMS/SMI approach." In Second IFIP International Conference on Wireless and Optical Communications Networks, 2005. WOCN 2005., pp. 603-606. IEEE, 2005.

23.	Mohamed I. AlHajri, Nazar T. Ali, and Raed M. Shubair. "A machine learning approach for the classification of indoor environments using RF signatures." In 2018 IEEE Global Conference on Signal and Information Processing (GlobalSIP), pp. 1060-1062. IEEE, 2018.

24.	E. M. Al-Ardi, R. M. Shubair, and M. E. Al-Mualla. "Performance evaluation of the LMS adaptive beamforming algorithm used in smart antenna systems." In 2003 46th Midwest Symposium on Circuits and Systems, vol. 1, pp. 432-435. IEEE, 2003.

25.	Satish R. Jondhale, Raed Shubair, Rekha P. Labade, Jaime Lloret, and Pramod R. Gunjal. "Application of supervised learning approach for target localization in wireless sensor network." In Handbook of Wireless Sensor Networks: Issues and Challenges in Current Scenario's, pp. 493-519. Springer, Cham, 2020.

26.	Raed Shubair, and Rashid Nuaimi. "Displaced sensor array for improved signal detection under grazing incidence conditions." Progress in Electromagnetics Research 79 (2008): 427-441.

27.	Raed M. Shubair, Abdulrahman S. Goian, Mohamed I. AlHajri, and Ahmed R. Kulaib. "A new technique for UCA-based DOA estimation of coherent signals." In 2016 16th Mediterranean Microwave Symposium (MMS), pp. 1-3. IEEE, 2016.

28.	M. I. AlHajri, R. M. Shubair, L. Weruaga, A. R. Kulaib, A. Goian, M. Darweesh, and R. AlMemari. "Hybrid method for enhanced detection of coherent signals using circular antenna arrays." In 2015 IEEE International Symposium on Antennas and Propagation \& USNC/URSI National Radio Science Meeting, pp. 1810-1811. IEEE, 2015.

29.	WafaNjima, Marwa Chafii, ArseniaChorti, Raed M. Shubair, and H. Vincent Poor. "Indoor localization using data augmentation via selective generative adversarial networks." IEEE Access 9 (2021): 98337-98347.

30.	Raed M. Shubair. "Improved smart antenna design using displaced sensor array configuration." Applied Computational Electromagnetics Society Journal 22, no. 1 (2007): 83.

31.	R. M. Shubair, and A. Merri. "A convergence study of adaptive beamforming algorithms used in smart antenna systems." In 11th International Symposium on Antenna Technology and Applied Electromagnetics [ANTEM 2005], pp. 1-5. IEEE, 2005.

32.	E. M. Ardi, , R. M. Shubair, and M. E. Mualla. "Adaptive beamforming arrays for smart antenna systems: A comprehensive performance study." In IEEE Antennas and Propagation Society Symposium, 2004., vol. 3, pp. 2651-2654. IEEE, 2004.

33.	Mohamed Ibrahim Alhajri, N. T. Ali, and R. M. Shubair. "2.4 ghz indoor channel measurements." IEEE Dataport (2018).

34.	Raed M. Shubair, and Hadeel Elayan. "Enhanced WSN localization of moving nodes using a robust hybrid TDOA-PF approach." In 2015 11th International Conference on Innovations in Information Technology (IIT), pp. 122-127. IEEE, 2015.

35.	M. I. AlHajri, N. T. Ali, and R. M. Shubair. "2.4 ghz indoor channel measurements data set." UCI Machine Learning Repository (2018).

36.	M. I. AlHajri, N. T. Ali, and R. M. Shubair. "2.4 ghz indoor channel measurements data set.” UCI Machine Learning Repository, 2018."

37.	WafaNjima, Marwa Chafii, and Raed M. Shubair. "Gan based data augmentation for indoor localization using labeled and unlabeled data." In 2021 International Balkan Conference on Communications and Networking (BalkanCom), pp. 36-39. IEEE, 2021.

38.	Mohamed I. AlHajri, Nazar T. Ali, and Raed M. Shubair. "A cascaded machine learning approach for indoor classification and localization using adaptive feature selection." AI for Emerging Verticals: Human-robot computing, sensing and networking (2020): 205.

39.	Mohamed I. AlHajri, Raed M. Shubair, and Marwa Chafii. "Indoor Localization Under Limited Measurements: A Cross-Environment Joint Semi-Supervised and Transfer Learning Approach." In 2021 IEEE 22nd International Workshop on Signal Processing Advances in Wireless Communications (SPAWC), pp. 266-270. IEEE, 2021.

40.	Raed M. Shubair and Hadeel Elayan. "In vivo wireless body communications: State-of-the-art and future directions." In 2015 Loughborough Antennas \& Propagation Conference (LAPC), pp. 1-5. IEEE, 2015.

41.	Hadeel Elayan, Raed M. Shubair, and Asimina Kiourti. "Wireless sensors for medical applications: Current status and future challenges." In 2017 11th European Conference on Antennas and Propagation (EUCAP), pp. 2478-2482. IEEE, 2017.

42.	Raed M. Shubair and Hadeel Elayan. "In vivo wireless body communications: State-of-the-art and future directions." In 2015 Loughborough Antennas \& Propagation Conference (LAPC), pp. 1-5. IEEE, 2015.

43.	Hadeel Elayan, Raed M. Shubair, and Asimina Kiourti. "Wireless sensors for medical applications: Current status and future challenges." In 2017 11th European Conference on Antennas and Propagation (EUCAP), pp. 2478-2482. IEEE, 2017.

44.	Hadeel Elayan, Raed M. Shubair, Josep Miquel Jornet, and Raj Mittra. "Multi-layer intrabody terahertz wave propagation model for nanobiosensing applications." Nano communication networks 14 (2017): 9-15.

45.	Hadeel Elayan, Pedram Johari, Raed M. Shubair, and Josep Miquel Jornet. "Photothermal modeling and analysis of intrabody terahertz nanoscale communication." IEEE transactions on nanobioscience 16, no. 8 (2017): 755-763.

46.	Rui Zhang, Ke Yang, Akram Alomainy, Qammer H. Abbasi, Khalid Qaraqe, and Raed M. Shubair. "Modelling of the terahertz communication channel for in-vivo nano-networks in the presence of noise." In 2016 16th Mediterranean Microwave Symposium (MMS), pp. 1-4. IEEE, 2016.

47.	Samar Elmeadawy and Raed M. Shubair. "6G wireless communications: Future technologies and research challenges." In 2019 international conference on electrical and computing technologies and applications (ICECTA), pp. 1-5. IEEE, 2019.

48.	Maryam AlNabooda, Raed M. Shubair, Nadeen R. Rishani, and GhadahAldabbagh. "Terahertz spectroscopy and imaging for the detection and identification of illicit drugs." 2017 Sensors networks smart and emerging technologies (SENSET) (2017): 1-4.

49.	Hadeel Elayan, Raed M. Shubair, and Asimina Kiourti. "On graphene-based THz plasmonic nano-antennas." In 2016 16th mediterranean microwave symposium (MMS), pp. 1-3. IEEE, 2016.

50.	Hadeel Elayan, Cesare Stefanini, Raed M. Shubair, and Josep Miquel Jornet. "End-to-end noise model for intra-body terahertz nanoscale communication." IEEE transactions on nanobioscience 17, no. 4 (2018): 464-473.

51.	Hadeel Elayan, Raed M. Shubair, Akram Alomainy, and Ke Yang. "In-vivo terahertz em channel characterization for nano-communications in wbans." In 2016 IEEE International Symposium on Antennas and Propagation (APSURSI), pp. 979-980. IEEE, 2016.

52.	Hadeel Elayan, Raed M. Shubair, and Josep M. Jornet. "Bio-electromagnetic thz propagation modeling for in-vivo wireless nanosensor networks." In 2017 11th European Conference on Antennas and Propagation (EuCAP), pp. 426-430. IEEE, 2017.

53.	Hadeel Elayan, and Raed M. Shubair. "On channel characterization in human body communication for medical monitoring systems." In 2016 17th International Symposium on Antenna Technology and Applied Electromagnetics (ANTEM), pp. 1-2. IEEE, 2016.

54.	Hadeel Elayan, Raed M. Shubair, and Nawaf Almoosa. "In vivo communication in wireless body area networks." In Information Innovation Technology in Smart Cities, pp. 273-287. Springer, Singapore, 2018.

55.	Hadeel Elayan, Raed M. Shubair, and Josep M. Jornet. "Characterising THz propagation and intrabody thermal absorption in iWNSNs." IET Microwaves, Antennas \& Propagation 12, no. 4 (2018): 525-532.

56.	Dana Bazazeh, Raed M. Shubair, and Wasim Q. Malik. "Biomarker discovery and validation for Parkinson's Disease: A machine learning approach." In 2016 International Conference on Bio-engineering for Smart Technologies (BioSMART), pp. 1-6. IEEE, 2016.

57.	Mayar Lotfy, Raed M. Shubair, Nassir Navab, and Shadi Albarqouni. "Investigation of focal loss in deep learning models for femur fractures classification." In 2019 International Conference on Electrical and Computing Technologies and Applications (ICECTA), pp. 1-4. IEEE, 2019.

58.	S. Elmeadawy, and R. M. Shubair. "Enabling technologies for 6G future wireless communications: Opportunities and challenges. arXiv 2020." arXiv preprint arXiv:2002.06068.

59.	Hadeel Elayan, Cesare Stefanini, Raed M. Shubair, and Josep M. Jornet. "Stochastic noise model for intra-body terahertz nanoscale communication." In Proceedings of the 5th ACM International Conference on Nanoscale Computing and Communication, pp. 1-6. 2018.

60.	Hadeel Elayan, Hadeel, and Raed M. Shubair. "Towards an Intelligent Deployment of Wireless Sensor Networks." In Information Innovation Technology in Smart Cities, pp. 235-250. Springer, Singapore, 2018.

61.	Abdul Karim Gizzini, Marwa Chafii, Ahmad Nimr, Raed M. Shubair, and Gerhard Fettweis. "Cnn aided weighted interpolation for channel estimation in vehicular communications." IEEE Transactions on Vehicular Technology 70, no. 12 (2021): 12796-12811.

62.	Nishtha Chopra, Mike Phipott, Akram Alomainy, Qammer H. Abbasi, Khalid Qaraqe, and Raed M. Shubair. "THz time domain characterization of human skin tissue for nano-electromagnetic communication." In 2016 16th Mediterranean Microwave Symposium (MMS), pp. 1-3. IEEE, 2016.

63.	Hadeel Elayan, Raed M. Shubair, Josep M. Jornet, Asimina Kiourti, and Raj Mittra. "Graphene-Based Spiral Nanoantenna for Intrabody Communication at Terahertz." In 2018 IEEE International Symposium on Antennas and Propagation \& USNC/URSI National Radio Science Meeting, pp. 799-800. IEEE, 2018.

64.	Taki Hasan Rafi, Raed M. Shubair, Faisal Farhan, Md Ziaul Hoque, and Farhan Mohd Quayyum. "Recent Advances in Computer-Aided Medical Diagnosis Using Machine Learning Algorithms with Optimization Techniques." IEEE Access (2021).

65.	Abdul Karim Gizzini, Marwa Chafii, Shahab Ehsanfar, and Raed M. Shubair. "Temporal Averaging LSTM-based Channel Estimation Scheme for IEEE 802.11 p Standard." arXiv preprint arXiv:2106.04829 (2021).

66.	Menna El Shorbagy, Raed M. Shubair, Mohamed I. AlHajri, and Nazih Khaddaj Mallat. "On the design of millimetre-wave antennas for 5G." In 2016 16th Mediterranean Microwave Symposium (MMS), pp. 1-4. IEEE, 2016.

67.	Ahmed A. Ibrahim,  JanMachac, and Raed M. Shubair. "Compact UWB MIMO antenna with pattern diversity and band rejection characteristics." Microwave and Optical Technology Letters 59, no. 6 (2017): 1460-1464.

68.	M. Saeed Khan, A-D. Capobianco, Sajid M. Asif, Adnan Iftikhar, Benjamin D. Braaten, and Raed M. Shubair. "A pattern reconfigurable printed patch antenna." In 2016 IEEE International Symposium on Antennas and Propagation (APSURSI), pp. 2149-2150. IEEE, 2016.

69.	M. Saeeed Khan, Adnan Iftikhar, Sajid M. Asif, Antonio‐Daniele Capobianco, and Benjamin D. Braaten. "A compact four elements UWB MIMO antenna with on‐demand WLAN rejection." Microwave and Optical Technology Letters 58, no. 2 (2016): 270-276.

70.	Muhammad Saeed Khan, Adnan Iftikhar, Antonio‐Daniele Capobianco, Raed M. Shubair, and Bilal Ijaz. "Pattern and frequency reconfiguration of patch antenna using PIN diodes." Microwave and Optical Technology Letters 59, no. 9 (2017): 2180-2185.

71.	Muhammad Saeed Khan, Adnan Iftikhar, Antonio‐Daniele Capobianco, Raed M. Shubair, and Bilal Ijaz. "Pattern and frequency reconfiguration of patch antenna using PIN diodes." Microwave and Optical Technology Letters 59, no. 9 (2017): 2180-2185.

72.	Muhammad Saeed Khan, Adnan Iftikhar, Raed M. Shubair, Antonio-D. Capobianco, Benjamin D. Braaten, and Dimitris E. Anagnostou. "Eight-element compact UWB-MIMO/diversity antenna with WLAN band rejection for 3G/4G/5G communications." IEEE Open Journal of Antennas and Propagation 1 (2020): 196-206.

73.	Amjad Omar, and Raed Shubair. "UWB coplanar waveguide-fed-coplanar strips spiral antenna." In 2016 10th European Conference on Antennas and Propagation (EuCAP), pp. 1-2. IEEE, 2016.

74.	Malak Y. ElSalamouny, and Raed M. Shubair. "Novel design of compact low-profile multi-band microstrip antennas for medical applications." In 2015 loughborough antennas \& propagation conference (LAPC), pp. 1-4. IEEE, 2015.

75.	Raed M. Shubair, Amna M. AlShamsi, Kinda Khalaf, and Asimina Kiourti. "Novel miniature wearable microstrip antennas for ISM-band biomedical telemetry." In 2015 Loughborough Antennas \& Propagation Conference (LAPC), pp. 1-4. IEEE, 2015.

76.	Ala Eldin Omer, George Shaker, Safieddin Safavi-Naeini, Georges Alquié, Frédérique Deshours, Hamid Kokabi, and Raed M. Shubair. "Non-invasive real-time monitoring of glucose level using novel microwave biosensor based on triple-pole CSRR." IEEE Transactions on Biomedical Circuits and Systems 14, no. 6 (2020): 1407-1420.

77.	Muhammad S. Khan, Syed A. Naqvi, Adnan Iftikhar, Sajid M. Asif, Adnan Fida, and Raed M. Shubair. "A WLAN band‐notched compact four element UWB MIMO antenna." International Journal of RF and Microwave Computer‐Aided Engineering 30, no. 9 (2020): e22282.

78.	Saad Alharbi, Raed M. Shubair, and Asimina Kiourti. "Flexible antennas for wearable applications: Recent advances and design challenges." (2018): 484-3.

79.	M. S. Khan, F. Rigobello, Bilal Ijaz, E. Autizi, A. D. Capobianco, R. Shubair, and S. A. Khan. "Compact 3‐D eight elements UWB‐MIMO array." Microwave and Optical Technology Letters 60, no. 8 (2018): 1967-1971.

80.	R. Karli, H. Ammor, R. M. Shubair, M. I. AlHajri, and A. Hakam. "Miniature Planar Ultra-Wide-Band Microstrip Patch Antenna for Breast Cancer Detection." Skin 1 (2016): 39.

81.	Mohammed S Al-Basheir, Raed M Shubai, and Sami M. Sharif. "Measurements and analysis for signal attenuation through date palm trees at 2.1 GHz frequency." (2006).

82.	Ala Eldin Omer, George Shaker, Safieddin Safavi-Naeini, Kieu Ngo, Raed M. Shubair, Georges Alquié, Frédérique Deshours, and Hamid Kokabi. "Multiple-cell microfluidic dielectric resonator for liquid sensing applications." IEEE Sensors Journal 21, no. 5 (2020): 6094-6104.

83.	Muhammad Saeed Khan, Adnan Iftikhar, Raed M. Shubair, Antonio-Daniele Capobianco, Sajid Mehmood Asif, Benjamin D. Braaten, and Dimitris E. Anagnostou. "Ultra-compact reconfigurable band reject UWB MIMO antenna with four radiators." Electronics 9, no. 4 (2020): 584.

84.	Amjad Omar, Maram Rashad, Maryam Al-Mulla, Hussain Attia, Shaimaa Naser, Nihad Dib, and Raed M. Shubair. "Compact design of UWB CPW-fed-patch antenna using the superformula." In 2016 5th International Conference on Electronic Devices, Systems and Applications (ICEDSA), pp. 1-4. IEEE, 2016.

85.	Muhammad S. Khan, Adnan Iftikhar, Raed M. Shubair, Antonio D. Capobianco, Benjamin D. Braaten, and Dimitris E. Anagnostou. "A four element, planar, compact UWB MIMO antenna with WLAN band rejection capabilities." Microwave and Optical Technology Letters 62, no. 10 (2020): 3124-3131.

86.	Ahmed A. Ibrahim, and Raed M. Shubair. "Reconfigurable band-notched UWB antenna for cognitive radio applications." In 2016 16th Mediterranean Microwave Symposium (MMS), pp. 1-4. IEEE, 2016.

87.	Hari Shankar Singh, SachinKalraiya, Manoj Kumar Meshram, and Raed M. Shubair. "Metamaterial inspired CPW‐fed compact antenna for ultrawide band applications." International Journal of RF and Microwave Computer‐Aided Engineering 29, no. 8 (2019): e21768.

88.	Omar Masood Khan, Qamar Ul Islam, Raed M. Shubair, and Asimina Kiourti. "Novel multiband Flamenco fractal antenna for wearable WBAN off-body communication applications." In 2018 International Applied Computational Electromagnetics Society Symposium (ACES), pp. 1-2. IEEE, 2018.

89.	Raed M. Shubair, Amer Salah, and Alaa K. Abbas. "Novel implantable miniaturized circular microstrip antenna for biomedical telemetry." In 2015 IEEE International Symposium on Antennas and Propagation \& USNC/URSI National Radio Science Meeting, pp. 947-948. IEEE, 2015.

90.	Sandip Ghosal, Arijit De, Ajay Chakrabarty, and Raed M. Shubair. "Characteristic mode analysis of slot loading in microstrip patch antenna." In 2018 IEEE International Symposium on Antennas and Propagation \& USNC/URSI National Radio Science Meeting, pp. 1523-1524. IEEE, 2018.

91.	Yazan Al-Alem, Ahmed A. Kishk, and Raed M. Shubair. "Enhanced wireless interchip communication performance using symmetrical layers and soft/hard surface concepts." IEEE Transactions on Microwave Theory and Techniques 68, no. 1 (2019): 39-50.

92.	Yazan Al-Alem, Ahmed A. Kishk, and Raed M. Shubair. "One-to-two wireless interchip communication link." IEEE Antennas and Wireless Propagation Letters 18, no. 11 (2019): 2375-2378.

93.	Yazan Al-Alem, Raed M. Shubair, and Ahmed Kishk. "Efficient on-chip antenna design based on symmetrical layers for multipath interference cancellation." In 2016 16th Mediterranean Microwave Symposium (MMS), pp. 1-3. IEEE, 2016.

94.	Nadeen R. Rishani, Raed M. Shubair, and GhadahAldabbagh. "On the design of wearable and epidermal antennas for emerging medical applications." In 2017 Sensors Networks Smart and Emerging Technologies (SENSET), pp. 1-4. IEEE, 2017.

95.	Asimina Kiourti, and Raed M. Shubair. "Implantable and ingestible sensors for wireless physiological monitoring: a review." In 2017 IEEE International Symposium on Antennas and Propagation \& USNC/URSI National Radio Science Meeting, pp. 1677-1678. IEEE, 2017.

96.	Yazan Al-Alem, Raed M. Shubair, and Ahmed Kishk. "Clock jitter correction circuit for high speed clock signals using delay units a nd time selection window." In 2016 16th Mediterranean Microwave Symposium (MMS), pp. 1-3. IEEE, 2016.

97.	Melissa Eugenia Diago-Mosquera, Alejandro Aragón-Zavala, Fidel Alejandro Rodríguez-Corbo, Mikel Celaya-Echarri, Raed M. Shubair, and Leyre Azpilicueta. "Tuning Selection Impact on Kriging-Aided In-Building Path Loss Modeling." IEEE Antennas and Wireless Propagation Letters 21, no. 1 (2021): 84-88.

98.	Mikel Celaya-Echarri, Leyre Azpilicueta, Fidel Alejandro Rodríguez-Corbo, Peio Lopez-Iturri, Victoria Ramos, Mohammad Alibakhshikenari, Raed M. Shubair, and Francisco Falcone. "Towards Environmental RF-EMF Assessment of mmWave High-Node Density Complex Heterogeneous Environments." Sensors 21, no. 24 (2021): 8419.

99.	Yazan Al-Alem, Ahmed A. Kishk, and Raed Shubair. "Wireless chip to chip communication link budget enhancement using hard/soft surfaces." In 2018 IEEE Global Conference on Signal and Information Processing (GlobalSIP), pp. 1013-1014. IEEE, 2018.

100.	Yazan Al-Alem, Yazan, Ahmed A. Kishk, and Raed M. Shubair. "Employing EBG in Wireless Inter-chip Communication Links: Design and Performance." In 2020 IEEE International Symposium on Antennas and Propagation and North American Radio Science Meeting, pp. 1303-1304. IEEE, 2020.

101.	Y. Lu, X. Zheng, “6G: A survey on technologies, scenarios, challenges, and the related issues,” Journal of Industrial Information Integration, vol. 19, Sept. 2020.
102.	W. Anani, A. Ouda and A. Hamou, "A Survey of Wireless Communications for IoT Echo-Systems," 2019 IEEE Canadian Conference of Electrical and Computer Engineering (CCECE), 2019, pp. 1-6.

103.	M. Gupta, R. K. Jha and S. Jain, "Tactile based Intelligence Touch Technology in IoT configured WCN in B5G/6G-A Survey," in IEEE Access.

104.	A. Yang, X. Yang, J. Chang, B. Bai, F. Kong and Q. Ran, "Research on a Fusion Scheme of Cellular Network and Wireless Sensor for Cyber Physical Social Systems," in IEEE Access, vol. 6, pp. 18786-18794, 2018.

105.	S.  Balaji, “IoT Technology, Applications and Challenges: A Contemporary Survey,” Wireless Personal Communications, vol. 108, pp. 363–388, April 2019.

106.	L. Babun et al., “A survey on IoT platforms: Communication, security, and privacy perspectives,” Computer Networks, vol. 192, June 2021.

107.	Q. Qi, X. Chen, C. Zhong and Z. Zhang, "Integration of Energy, Computation and Communication in 6G Cellular Internet of Things," in IEEE Communications Letters, vol. 24, no. 6, pp. 1333-1337, June 2020.

108.	S. Li et al., “5G Internet of Things: A survey,” Journal of Industrial Information Integration, vol. 10, pp. 1-9, June 2018.

109.	F. Al-Ogaili and R. M. Shubair, "Millimeter-wave mobile communications for 5G: Challenges and opportunities," 2016 IEEE International Symposium on Antennas and Propagation (APSURSI), 2016, pp. 1003-1004.

110.	C. R. Storck and F. Duarte-Figueiredo, "A Survey of 5G Technology Evolution, Standards, and Infrastructure Associated With Vehicle-to-Everything Communications by Internet of Vehicles," in IEEE Access, vol. 8, pp. 117593-117614, 2020.

111.	A. Dogra, R. K. Jha and S. Jain, "A Survey on Beyond 5G Network With the Advent of 6G: Architecture and Emerging Technologies," in IEEE Access, vol. 9, pp. 67512-67547, 2021.

112.	C. D. Alwis et al., "Survey on 6G Frontiers: Trends, Applications, Requirements, Technologies and Future Research," in IEEE Open Journal of the Communications Society, vol. 2, pp. 836-886, 2021.

113.	M. Alsabah et al., "6G Wireless Communications Networks: A Comprehensive Survey," in IEEE Access, vol. 9, pp. 148191-148243, 2021.

114.	L. Qiao et al., “A survey on 5G/6G, AI, and Robotics,” Computers \& Electrical Engineering, vol. 95, Oct. 2021.

115.	X. You, et al., “Towards 6G wireless communication networks: vision, enabling technologies, and new paradigm shifts,” Science China Information Sciences, vol. 64, Nov. 2020.

116.	H. Elayan, O. Amin, B. Shihada, R. M. Shubair and M. -S. Alouini, "Terahertz Band: The Last Piece of RF Spectrum Puzzle for Communication Systems," in IEEE Open Journal of the Communications Society, vol. 1, pp. 1-32, 2020.

117.	H. Elayan, O. Amin, R. M. Shubair and M. -S. Alouini, "Terahertz communication: The opportunities of wireless technology beyond 5G," 2018 International Conference on Advanced Communication Technologies and Networking (CommNet), 2018, pp. 1-5.

118.	C. Huang et al., "Holographic MIMO Surfaces for 6G Wireless Networks: Opportunities, Challenges, and Trends," in IEEE Wireless Communications, vol. 27, no. 5, pp. 118-125, Oct. 2020.

119.	Q. Wu, S. Zhang, B. Zheng, C. You and R. Zhang, "Intelligent Reflecting Surface-Aided Wireless Communications: A Tutorial," in IEEE Transactions on Communications, vol. 69, no. 5, pp. 3313-3351, May 2021.

120.	R. Liu, Q. Wu, M. Di Renzo and Y. Yuan, "A Path to Smart Radio Environments: An Industrial Viewpoint on Reconfigurable Intelligent Surfaces," in IEEE Wireless Communications, vol. 29, no. 1, pp. 202-208, Feb. 2022.

121.	Y. -C. Liang, Q. Zhang, E. G. Larsson and G. Y. Li, "Symbiotic Radio: Cognitive Backscattering Communications for Future Wireless Networks," in IEEE Transactions on Cognitive Communications and Networking, vol. 6, no. 4, pp. 1242-1255, Dec. 2020.

122.	S. Chen, J. Zhang, Y. Jin and B. Ai, "Wireless powered IoE for 6G: Massive access meets scalable cell-free massive MIMO," in China Communications, vol. 17, no. 12, pp. 92-109, Dec. 2020.

123.	X. Zhu and C. Jiang, "Integrated Satellite-Terrestrial Networks Toward 6G: Architectures, Applications, and Challenges," in IEEE Internet of Things Journal, vol. 9, no. 1, pp. 437-461, Jan. 2022.

124.	Z. Wang, Z. Zhou, H. Zhang, G. Zhang, H. Ding and A. Farouk, "AI-Based Cloud-Edge-Device Collaboration in 6G Space-Air-Ground Integrated Power IoT," in IEEE Wireless Communications, vol. 29, no. 1, pp. 16-23, Feb. 2022.

125.	S. Zhang and R. Zhang, "Intelligent Reflecting Surface Aided Multi-User Communication: Capacity Region and Deployment Strategy," in IEEE Transactions on Communications, vol. 69, no. 9, pp. 5790-5806, Sept. 2021.

126.	C. You, B. Zheng and R. Zhang, "Channel Estimation and Passive Beamforming for Intelligent Reflecting Surface: Discrete Phase Shift and Progressive Refinement," in IEEE Journal on Selected Areas in Communications, vol. 38, no. 11, pp. 2604-2620, Nov. 2020.

127.	S. Rajoria, A. Trivedi, and W. W. Godfrey, “A comprehensive survey: Small cell meets massive MIMO,” Physical Communication, vol. 26, pp. 40-49, February 2018.

128.	M. H. Alsharif, R. Nordin, M. M. Shakir, “Small Cells Integration with the Macro-Cell Under LTE Cellular Networks and Potential Extension for 5G,” Journal of Electrical Engineering and Technology, vol. 14, pp. 2455–2465, April 2019.

129.	A. Rejeb et al., “Internet of Things research in supply chain management and logistics: A bibliometric analysis,” Internet of Things, vol. 12, Dec. 2020.

130.	A. Vakaloudis and C. O’Leary, "A framework for rapid integration of IoT Systems with industrial environments," 2019 IEEE 5th World Forum on Internet of Things (WF-IoT), 2019, pp. 601-605.

131.	F. Tlili et al., “Design and architecture of smart belt for real time posture monitoring,” Internet of Things, vol 27, Mar. 2022.

132.	M. N. Bhuiyan, M. M. Rahman, M. M. Billah and D. Saha, "Internet of Things (IoT): A Review of Its Enabling Technologies in Healthcare Applications, Standards Protocols, Security, and Market Opportunities," in IEEE Internet of Things Journal, vol. 8, no. 13, pp. 10474-10498, 1 July1, 2021.

133.	W. C. Tchuitcheu et al., “Internet of smart-cameras for traffic lights optimization in smart cities,” Internet of Things, vol. 11, Sept. 2020.

134.	A. Kirimtat, O. Krejcar, A. Kertesz and M. F. Tasgetiren, "Future Trends and Current State of Smart City Concepts: A Survey," in IEEE Access, vol. 8, pp. 86448-86467, 2020.

135.	V. Williams, S. Terence J. and J. Immaculate, "Survey on Internet of Things based Smart Home," 2019 International Conference on Intelligent Sustainable Systems (ICISS), 2019, pp. 460-464.

136.	I. Charania and X. Li, “Smart farming: Agriculture's shift from a labor intensive to technology native industry,” Internet of Things, vol. 9, Mar. 2020.

137.	M. S. Farooq, S. Riaz, A. Abid, K. Abid and M. A. Naeem, "A Survey on the Role of IoT in Agriculture for the Implementation of Smart Farming," in IEEE Access, vol. 7, pp. 156237-156271.

138.	M. H. Cintuglu, O. A. Mohammed, K. Akkaya and A. S. Uluagac, "A Survey on Smart Grid Cyber-Physical System Testbeds," in IEEE Communications Surveys \& Tutorials, vol. 19, no. 1, pp. 446-464, Firstquarter 2017.

139.	J. Zhang and D. Tao, "Empowering Things With Intelligence: A Survey of the Progress, Challenges, and Opportunities in Artificial Intelligence of Things," in IEEE Internet of Things Journal, vol. 8, no. 10, pp. 7789-7817, 15 May15, 2021.

140.	S. Messaoud et al., “A survey on machine learning in Internet of Things: Algorithms, strategies, and applications,” Internet of Things, vol. 12, Dec. 2020.

141.	M. A. Al-Garadi, A. Mohamed, A. K. Al-Ali, X. Du, I. Ali and M. Guizani, "A Survey of Machine and Deep Learning Methods for Internet of Things (IoT) Security," in IEEE Communications Surveys \& Tutorials, vol. 22, no. 3, pp. 1646-1685, thirdquarter 2020.

142.	Y. Xie, Y. Hu, Y. Chen, Y. Liu and G. Shou, "A Video Analytics-Based Intelligent Indoor Positioning System Using Edge Computing For IoT," 2018 International Conference on Cyber-Enabled Distributed Computing and Knowledge Discovery (CyberC), 2018, pp. 118-1187.

143.	A. Alam, I. Ullah and Y. -K. Lee, "Video Big Data Analytics in the Cloud: A Reference Architecture, Survey, Opportunities, and Open Research Issues," in IEEE Access, vol. 8, pp. 152377-152422, 2020.

144.	S. Y. Jang, Y. Lee, B. Shin and D. Lee, "Application-Aware IoT Camera Virtualization for Video Analytics Edge Computing," 2018 IEEE/ACM Symposium on Edge Computing (SEC), 2018, pp. 132-144.

145.	H. Razalli, M. H. Alkawaz and A. S. Suhemi, "Smart IOT Surveillance Multi-Camera Monitoring System," 2019 IEEE 7th Conference on Systems, Process and Control (ICSPC), 2019, pp. 167-171.

146.	G. Gagliardi et al., “An Internet of Things Solution for Smart Agriculture,” Agronomy, vol. 11, no. 11, Oct. 2021.

147.	S. Elmeadawy and R. M. Shubair, "6G Wireless Communications: Future Technologies and Research Challenges," 2019 International Conference on Electrical and Computing Technologies and Applications (ICECTA), 2019, pp. 1-5.

148.	X. Xie et al., "A Joint Optimization Framework for IRS-assisted Energy Self-sustainable IoT Networks," in IEEE Internet of Things Journal.

149.	G. Yu, X. Chen, C. Zhong, H. Lin and Z. Zhang, "Large Intelligent Reflecting Surface Enhanced Massive Access for B5G Cellular Internet of Things," 2020 IEEE 91st Vehicular Technology Conference (VTC2020-Spring), 2020, pp. 1-5.

150.	A. Mahmoud, S. Muhaidat, P. C. Sofotasios, I. Abualhaol, O. A. Dobre and H. Yanikomeroglu, "Intelligent Reflecting Surfaces Assisted UAV Communications for IoT Networks: Performance Analysis," in IEEE Transactions on Green Communications and Networking, vol. 5, no. 3, pp. 1029-1040, Sept. 2021.

151.	J. Wu and B. Shim, "Power Minimization of Intelligent Reflecting Surface-Aided Uplink IoT Networks," 2021 IEEE Wireless Communications and Networking Conference (WCNC), 2021, pp. 1-6.

152.	W. Hao et al., "Robust Design for Intelligent Reflecting Surface-Assisted MIMO-OFDMA Terahertz IoT Networks," in IEEE Internet of Things Journal, vol. 8, no. 16, pp. 13052-13064, 15 Aug.15, 2021.

153.	G. Yu, X. Chen, C. Zhong, D. W. Kwan Ng and Z. Zhang, "Design, Analysis, and Optimization of a Large Intelligent Reflecting Surface-Aided B5G Cellular Internet of Things," in IEEE Internet of Things Journal, vol. 7, no. 9, pp. 8902-8916, Sept. 2020.

154.	Z. Chu, P. Xiao, M. Shojafar, D. Mi, J. Mao and W. Hao, "Intelligent Reflecting Surface Assisted Mobile Edge Computing for Internet of Things," in IEEE Wireless Communications Letters, vol. 10, no. 3, pp. 619-623, Mar. 2021.

155.	X. Li, Q. Zhu and Y. Wang, "IRS-Assisted Crowd Spectrum Sensing in B5G Cellular IoT Networks," 2020 International Conference on Wireless Communications and Signal Processing (WCSP), 2020, pp. 761-765.

156.	Z. Chu, Z. Zhu, X. Li, F. Zhou, L. Zhen and N. Al-Dhahir, "Resource Allocation for IRS Assisted Wireless Powered FDMA IoT Networks," in IEEE Internet of Things Journal.

157.	F. C. Okogbaa et al., “Design and Application of Intelligent Reflecting Surface (IRS) for Beyond 5G Wireless Networks: A Review,” MDPI-Sensors, vol. 22, no. 7, March 2022, doi: 10.3390/s22072436.
158.	F. Guo, F. R. Yu, H. Zhang, X. Li, H. Ji and V. C. M. Leung, "Enabling Massive IoT Toward 6G: A Comprehensive Survey," in IEEE Internet of Things Journal, vol. 8, no. 15, pp. 11891-11915, Aug. 2021.

159.	S. Verma, S. Kaur, M. A. Khan and P. S. Sehdev, "Toward Green Communication in 6G-Enabled Massive Internet of Things," in IEEE Internet of Things Journal, vol. 8, no. 7, pp. 5408-5415, April 2021.

160.	T. Mir, L. Dai, Y. Yang, W. Shen and B. Wang, "Optimal FemtoCell Density for Maximizing Throughput in 5G Heterogeneous Networks under Outage Constraints," 2017 IEEE 86th Vehicular Technology Conference (VTC-Fall), Toronto, ON, Canada, 2017, pp. 1-5.

161.	N. Hassan and X. Fernando, "Interference Mitigation and Dynamic User Association for Load Balancing in Heterogeneous Networks," in IEEE Transactions on Vehicular Technology, vol. 68, no. 8, pp. 7578-7592, Aug. 2019.

162.	M. Mozaffari, W. Saad, M. Bennis and M. Debbah, "Optimal Transport Theory for Cell Association in UAV-Enabled Cellular Networks," in IEEE Communications Letters, vol. 21, no. 9, pp. 2053-2056, Sept. 2017.

163.	M. Mozaffari, W. Saad, M. Bennis and M. Debbah, "Performance Optimization for UAV-Enabled Wireless Communications under Flight Time Constraints," GLOBECOM 2017 - 2017 IEEE Global Communications Conference, 2017, pp. 1-6.

164.	Q. -V. Pham, L. B. Le, S. -H. Chung and W. -J. Hwang, "Mobile Edge Computing With Wireless Backhaul: Joint Task Offloading and Resource Allocation," in IEEE Access, vol. 7, pp. 16444-16459, 2019.

165.	W. Kim, "Dual Connectivity in Heterogeneous Small Cell Networks with mmWave Backhauls", in Mobile Information Systems, vol. 2016, pp. 1-14 pages, Oct. 2016.

166.	Z. Xiao, H. Liu, V. Havyarimana, T. Li, D. Wang, “Analytical Study on Multi-Tier 5G Heterogeneous Small Cell Networks: Coverage Performance and Energy Efficiency,” in Sensors, vol. 16, no. 11, Nov. 2016.

167.	W. Tang et al., "Wireless Communications With Reconfigurable Intelligent Surface: Path Loss Modeling and Experimental Measurement," in IEEE Transactions on Wireless Communications, vol. 20, no. 1, pp. 421-439, Jan. 2021.

168.	M. O. Al-Kadri, Y. Deng, A. Aijaz and A. Nallanathan, "Full-Duplex Small Cells for Next Generation Heterogeneous Cellular Networks: A Case Study of Outage and Rate Coverage Analysis," in IEEE Access, vol. 5, pp. 8025-8038, 2017.

169.	M. M. Fadoul, “Rate and Coverage Analysis in Multi-tier Heterogeneous Network Using Stochastic Geometry Approach,” in Ad Hoc Networks, vol.  98. 102038, Mar. 2020.

170.	A. Alshahrani, I. A. Elgendy, A. Muthanna, A. M. Alghamdi, A. Alshamrani, “Efficient Multi-Player Computation Offloading for VR Edge-Cloud Computing Systems,” Applied Sciences, vol. 10, no. 16, Aug. 2020.

171.	D. Xu, X. Yu and R. Schober, "Resource Allocation for Intelligent Reflecting Surface-Assisted Cognitive Radio Networks," 2020 IEEE 21st International Workshop on Signal Processing Advances in Wireless Communications (SPAWC), 2020, pp. 1-5.

172.	Q. Wu and R. Zhang, "Intelligent Reflecting Surface Enhanced Wireless Network via Joint Active and Passive Beamforming," in IEEE Transactions on Wireless Communications, vol. 18, no. 11, pp. 5394-5409, Nov. 2019.

173.	E. Björnson, Ö. Özdogan and E. G. Larsson, "Intelligent Reflecting Surface Versus Decode-and-Forward: How Large Surfaces are Needed to Beat Relaying?," in IEEE Wireless Communications Letters, vol. 9, no. 2, pp. 244-248, Feb. 2020.

174.	Z. Yang et al., "Energy-Efficient Wireless Communications With Distributed Reconfigurable Intelligent Surfaces," in IEEE Transactions on Wireless Communications, vol. 21, no. 1, pp. 665-679, Jan. 2022.

175.	Y. Cao, T. Lv and W. Ni, "Intelligent Reflecting Surface Aided Multi-User mmWave Communications for Coverage Enhancement," 2020 IEEE 31st Annual International Symposium on Personal, Indoor and Mobile Radio Communications, 2020, pp. 1-6.

176.	A. Al-Hilo et al., “RIS-Assisted UAV for Timely Data Collection in IoT Networks,” 2021, arXiv: 2103.17162v2. [Online]. Available: https://arxiv.org/abs/2103.17162. 
177.	M. TEKBAŞ, A. TOKTAŞ and G. ÇAKIR, "Design of a Dual Polarized mmWave Horn Antenna Using Decoupled Microstrip Line Feeder," 2020 International Conference on Electrical Engineering (ICEE), 2020, pp. 1-4.

178.	M. W. Akhtar et al., “The shift to 6G communications: vision and requirements,” Human-centric Computing and Information Sciences, vol. 10, Dec. 2020.

179.	B. Picano et al., “End-to-End Delay Bound for VR Services in 6G Terahertz Networks with Heterogeneous Traffic and Different Scheduling Policies,” Mathematics, vol. 9, no. 14, July 2021.

180.	S. Chen, Y. -C. Liang, S. Sun, S. Kang, W. Cheng and M. Peng, "Vision, Requirements, and Technology Trend of 6G: How to Tackle the Challenges of System Coverage, Capacity, User Data-Rate and Movement Speed," in IEEE Wireless Communications, vol. 27, no. 2, pp. 218-228, April 2020.

}

\end{document}